# The Effect of Different Wavelengths on Porous Silicon Formation Process

**Oday A. Abbas**

**Department of Physics**
**College of Sciences/ University of Kufa**

**Adel H. Omran**

**Department of Physics**
**College of Sciences/ University of Kufa**

**Abbas H. Rahim**

**Department of Physics**
**College of Sciences/ University of Kufa**

## *ABSTRACT*

Porous silicon layers (PS) have been prepared in this work via Photo-electrochemical etching process (PEC) of n-type silicon wafer of 0.8 $\Omega$.cm resistivity in hydrofluoric (HF) acid of 24.5 % concentration at different etching times (5 – 25min.). The irradiation has been achieved using Tungsten lamp with different wavelengths (450 nm, 535 nm and 700 nm). The morphological properties of these layers such as surface morphology, Porosity, layer thickness, and also the etching rate have been investigated using optical microscopy and the gravimetric method.

## *1. INTRODUCTION*

Bulk crystalline silicon has an indirect band-gap of 1.12 eV at room temperature. Due to the non-radiative transition mechanism of the indirect band-gap structure, the probability of efficient inter band radiative-recombination is extremely low for bulk silicon under normal circumstance. One way to overcome such a low intrinsic radiative recombination rate is to form porous silicon (PS) [1]. Porous silicon can be considered as a silicon crystal having a network of voids in it. The nanosized voids in the silicon bulk result in a sponge-like structure of pores and channels surrounded with a skeleton of crystalline silicon nanowires. The structural properties of porous silicon are fundamentally determined by the porosity and thickness of PS layer. These parameters depend on preparation conditions [2]. The first group which is described the formation of PS layer on silicon electrodes in hydrofluoric acid electrolytes under anodic bias in 1956 are they *Uhlir and Turner*. The discovery of a form of PS with photoluminescence efficiency much higher than that of bulk silicon has led to a flurry of research activity [3]. Many studies have shown that certain electrochemical and/or chemical etches of silicon produce PS material that photoluminesces in the visible region of the electromagnetic spectrum [4]. The remarkable luminescence of this material has generated interest because of the unique role that PS plays in modern electronics and optic-electronics technologies [5]. The PS material can be also produced by a photoelectrochemical etching process which gives advantage of the rectifying nature of the semiconductor/liquid junction; illumination reduces the net etch rate (hole current) at p-type silicon, while it increases the etch rate at n-type silicon [6]. In this research, we investigated the effects of incoherent light with different wavelengths during electrochemical etching process on the formation processes of PS layer and its morphological properties.

## 2. EXPERIMENT

Crystalline wafer of n-type Silicon with resistivity of 0.8 $\Omega$.cm, 508 $\mu$m thickness, and (111) orientation was used as starting substrate. The substrate was cut into rectangles with areas of 1 cm$^2$. The native oxide was cleaned in a mixture of HF and $H_2O$ (1:2). After chemical treatment, 0.1 μm-thick Al layers were deposited, by using an evaporation method, on the backsides of the wafer. Photoelectrochemical etching then performed in an ethanoic solution of 24.5% HF is obtained by 1 volume of ethanol and 1 volume of 49% wt. Hydrofluoric (HF) acid (the HF concentration in the ethanoic solution is given by $((1 \times 49\%)/(1+1) = 24.5\%)$) at room temperature by using a Pt electrode. Current of 30 mA/cm$^2$ was applied; Samples were illuminated in three ways using a halogen lamp (Philips Diachronic) with (450 nm) wavelength in the first and (535 nm) in the second, while in the third by (700 nm). The morphological properties: surface morphology, layer thickness, pore shape have been measured using high resolution optical microscope (OLUMPUS BH2) connected to a computer and a digital camera. We have used almost the same magnification power for studying the surface morphology which is (600 x).

The porosity has been measured using Gravimetric method and using the following equation [7]:

$$\gamma = \frac{M_1 - M_2}{M_1 - M_3} \dots\dots\dots\dots\dots\dots\dots\dots\dots\dots(1)$$

where $M_1$ (g) and $M_2$ (g) are the weights of silicon sample before and after etching process respectively. $M_3$ (g) is the weight of the silicon sample after removing of PS layer. The etching rate has been measured using the following equation [8]:

$$v = \frac{d}{t} \dots\dots\dots\dots\dots\dots\dots\dots\dots\dots\dots\dots(2)$$

where $v$ (μm/min) is the etching rate, $d$ (μm) is the layer thickness and $t$ (min.) is the etching time.

# 3. RESULTS AND DISCUSSION
## 3.1. Surface morphology

Figure (1) represents optical microscopy images (top-view) of PS layers prepared at different etching times (5-15 min.) with constant current density of 30 mA/cm$^2$. The irradiation has been achieved using blue light (450 nm) as shown in figure (1, a) and then with Green light (535 nm) as shown in figure (1, b), while with red light (700 nm) as shown in figure (1, c) on (0.8 Ω.cm) n-type silicon immersed in 24.5 % HF concentration.

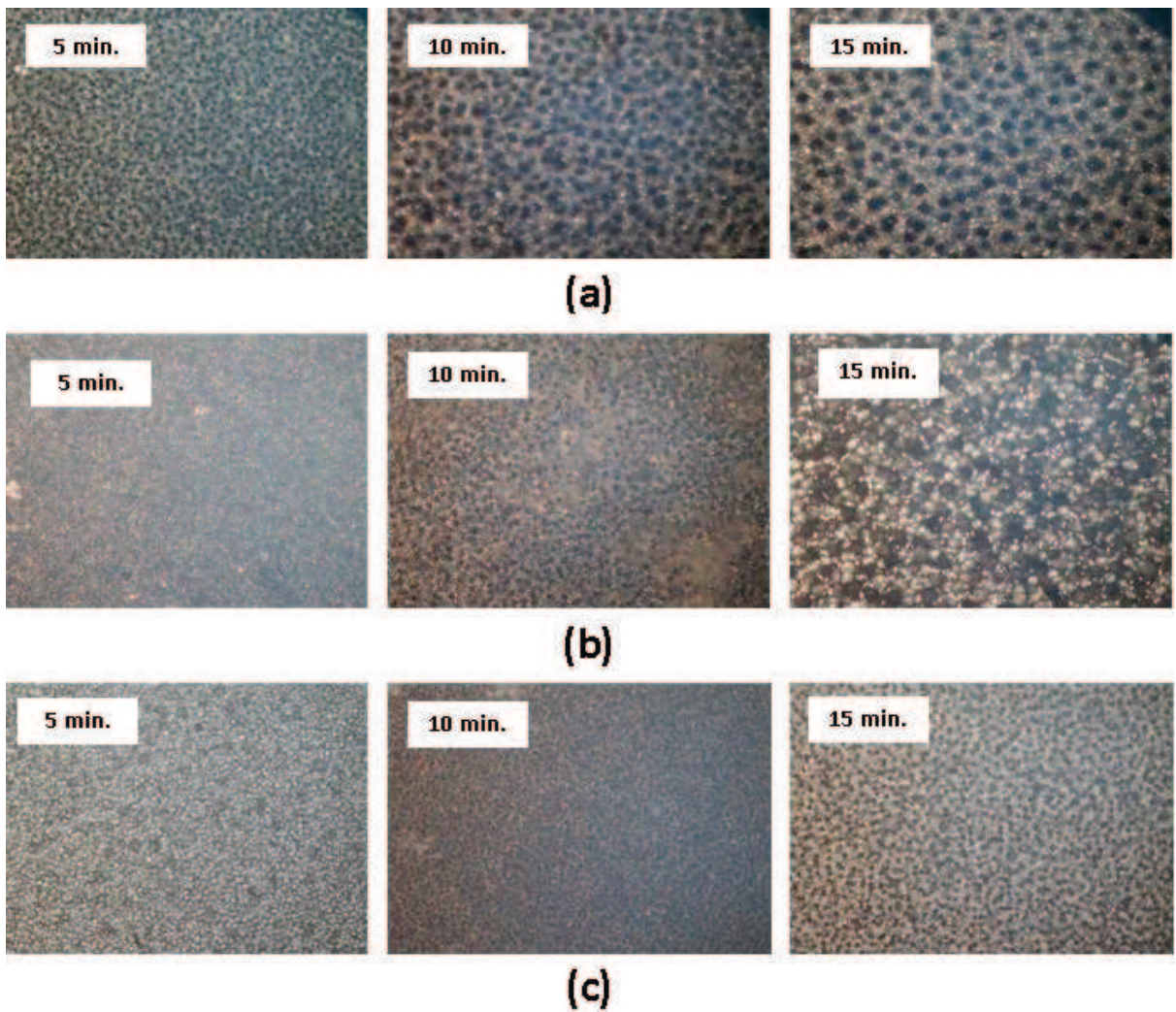

Figure (1); Optical microscopy images (top-view) of prepared PS layers.

From figure (1) we can arrive to several facts as follows:

1- The pore width of pores in PS layers have been prepared under 450 nm wavelength bigger than that of pores in PS layers prepared under 535 nm which is larger than that of pores in PS layers prepared under 700 nm. Here we interpret these results in terms of two competing processes; the first is the electrochemical etching process, which is driven by the potential applied to the silicon electrode and can occur only at regions within the silicon that are in good electrical contact with the substrate. This process is happen in all cases, while the second process is a purely photochemical corrosion that can happen even at electrically isolated regions within or on the surface of the PS material and this process has been very active at irradiation with 450 nm wavelength because it can be absorbed at the silicon surface which will promote the charge carriers (holes ) to cross the potential barrier at the interface between silicon and the HF solution (Schottky barrier), while in the other wavelengths (535 and 700 nm) the absorption happen at silicon substrate. Our results in good agreement with investigations of [9,10].

2- The distribution of pores in the PS layers was varied as shown in figure (1); we can note that the distribution of pores in prepared PS layers under 450 nm was homogenous, while in the other cases, the density of pores is varying from position to another. Depending on the reported facts by *Canham and Lehmann* [11,12] is that photochemical etching process for n-type silicon is homogenous while electrochemical etching process not homogenous, therefore we ascribe these results to photochemical etching effect which was the first process for formatting pores while in the other cases the electrochemical etching was the first process for formatting the pores.

3- The indelicacy of surface of PS layers which is as a function of porosity is decreasing with increasing of wavelength but in same time it increases with increasing of etching time.

This result has been ascribed to increasing of generated charge carrier (holes) which will increase the etching process and then formation of pores. To avoid the ambiguity, the formation process of PS layers can be summarized as follows: A silicon surface is known to be virtually inert against attack of HF acid [13]. If the silicon is under anodic bias, the reaction begins with the generation of holes, either thermally generated or photogenerated. The holes drift under the applied bias to the interface between silicon and solution, where they can react, and then formation PS layer is observed as long as the reaction is limited by the charge supply of the electrode. This condition is fulfilled for current density (J) below the critical current density ($J_{PS}$) [13,14].

## *3.2. Layer thickness*

Figure (2) shows the relation between thickness of PS layer and etching time of PS layers prepared under different wavelengths.

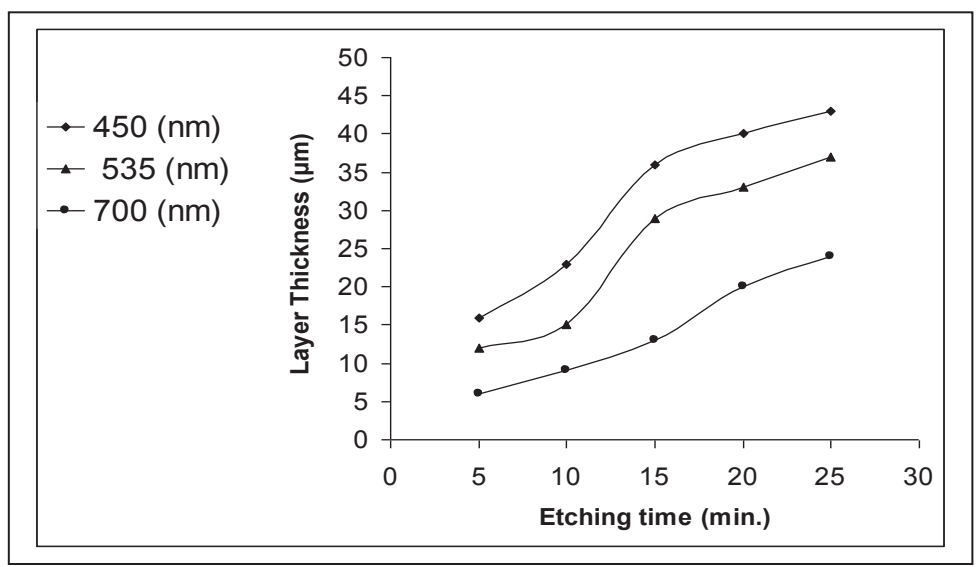

Figure (2); Relation between layer thickness and etching time of prepared PS layers.

As shown in previous figure the value of thickness of PS layer prepared under 450 nm larger than that of PS layers fabricated under 535 nm, which itself bigger than that which manufactured under 700 nm. We ascribed these results to fact that the photochemical etching process with 450 nm is more vigorous than that with (535 nm and 700 nm). Also we can note that the layer thickness increases with increasing of etching time in all cases. Our explanation for this result can be abstracted in the following manner, the increasing of etching time lead to increasing of required charge carriers (holes) for formation process and then increases this process which leading to increase of layer thickness.

## *3.3 Porosity*

Figure (3) illustrates the effect of etching time on value of porosity of PS layers prepared under different wavelengths. We observe that the value of porosity of prepared PS layers in all cases has been increased with increasing of etching time and decrease of used wavelength. This result can be explained upon fundament that the corrosion effect increases with increasing of etching time and decrease of applied wavelength and which lead to increasing porosity value.

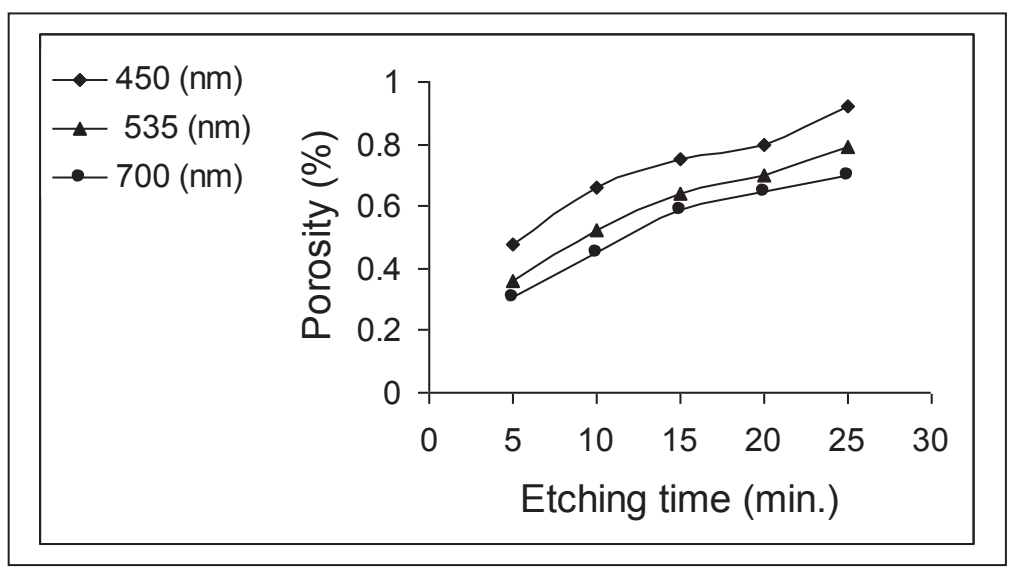

Figure (3); The effect of etching time on value of porosity of prepared PS layers.

## *3.4. Etching rate*

Figure (4) shows the relation between etching time and etching rate of PS layers prepared under different wavelength.

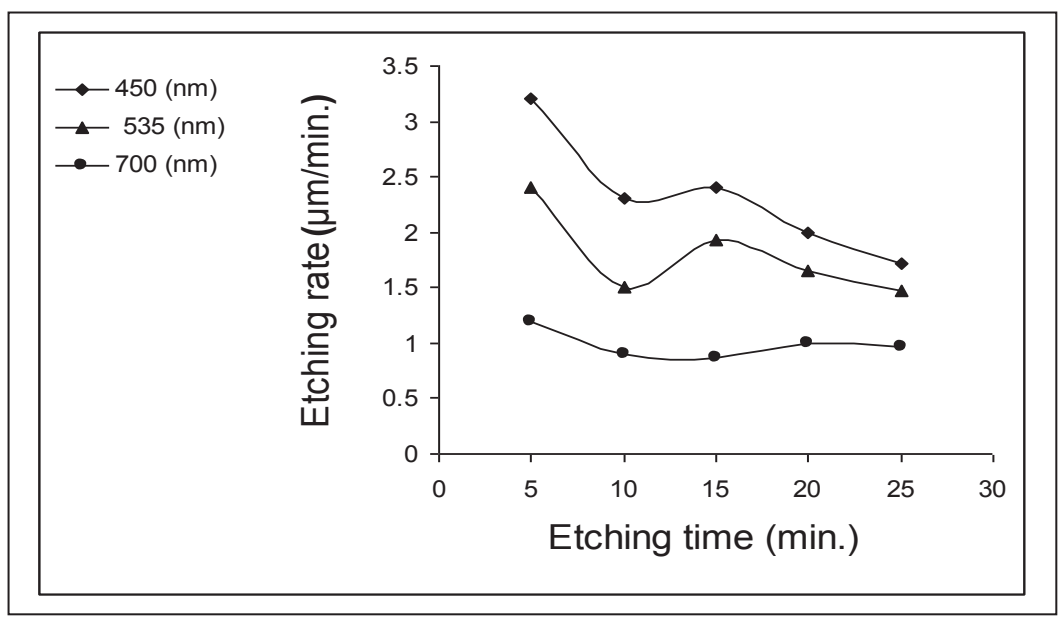

Figure (4); The relation between etching time and etching rate.

From this figure, we can confirm that the value of etching rate under short wavelengths bigger than that under long wavelengths. These results can be cleared with the following scenario: Generally, single-crystal silicon electrodes immersed in HF etching solutions display rectification properties analogous to Schottky junctions. The anodic current necessary for the corrosion reactant that forms PS layer corresponds to the reverse bias current for n-type silicon as in Schottky solar cells [4]. Porous silicon typically forms in a columnar structure. The electrochemical etching process that produces PS layer can occur only in regions where the silicon columns are in good electrical contact with the substrate.

Thus, as the fibers within the PS layer become thinner, the holes required to etch the material are less likely to make it from the substrate into these thinner regions.

In contrast, a photoetching process can be much more effective at etching the thinner regions. The short penetration depths of blue (450 nm) light in silicon (0.43 μm) and Green (535 nm) light in silicon (0.87 μm) [15] ensures that electron-hole pairs are generated close to the fiber/solution interface. It is known that in bulk silicon crystals, electronic holes generated close enough to the surface are capable of overcoming the built-in potential of the Schottky junction and crossing the interface. In the case of PS, this effect should become even more pronounced as the size of the silicon fibers becomes smaller than the width of the space-charge region. Thus the close proximity of holes to the surface and the small dimensions of the silicon fibers combine to enhance the photochemical etching process for short illumination wavelengths. Alternatively, red (700 nm) light has a penetration depth (8.4 μm) [15] greater than the thickness of the primarily PS layer (0-5 μm), and so red illumination is expected to result primarily in excitation of the substrate only.

# *4. CONCLUSIONS*

We can conclude from previous results and facts that the photoelectrochemical etching process is useful method for production of PS layers and also the formation process of PS layer and its morphological properties depends on the etching conditions (etching time and illumination (wavelength)). The prepared PS layers at 25 min. etching time and under 450 nm wavelength has high values of pore width, porosity, layer thickness and etching rate, while the prepared PS layers at 5 min. etching time and under 700 nm has low values of pore width, porosity, layer thickness and etching rate.

# 5. REFRENCES

# تأثير الاطوال الموجية المختلفة على عملية تشكُل السليكون المسامي

**عادل حبيب عمران**          **عدي اركان عباس**

**جامعة الكوفة/كلية العلوم/قسم الفيزياء**          **جامعة الكوفة/كلية العلوم/قسم الفيزياء**

**عباس حسن رحيم**

**جامعة الكوفة/كلية العلوم/قسم الفيزياء**

## الخلاصة

تم في هذا البحث تحضير طبقات السليكون المسامي بواسطة عملية القشط الضوئي الكهروكيميائي لشرائح سليكون ذات توصيلية كهربائية من نوع n وبمقاومية كهربائية نوعية مقدارها ($\Omega.cm$ 0.8) في حامض الهايدروفلوريك (HF) بتركيز 24.5 % عند أزمنة قشط مختلفة (.min 25 – 5). التشعيع انجز باستخدام اشعاع مصباح التنكستن باطوال موجية مختلفة (450 nm, 535 nm و 700 nm). تم دراسة الخصائص الطوبوغرافية لطبقات السليكون المسامي المحضرة مثل طوبوغرافية السطح, المسامية, سمك الطبقة وكذلك نسبة القشط باستخدام المجهر الضوئي و الطريقة الوزنية.